\documentclass[pra,amsmath,aps,10pt,superscriptaddress,letterpaper,tightenlines,showpacs,twocolumn]{revtex4}
\usepackage{bm}
\usepackage{epstopdf}
\usepackage{stmaryrd}
\usepackage{amsmath}
\usepackage{calligra}
\usepackage{mathtools}
\usepackage{amsfonts}
\usepackage{amssymb}
\usepackage{textcomp}
\usepackage{graphicx}
\usepackage{caption}
\usepackage[margin=15mm]{geometry}
\usepackage{yfonts}
\usepackage[breaklinks=true,colorlinks=true,linkcolor=blue,urlcolor=blue,citecolor=blue]{hyperref}
\usepackage[lofdepth,lotdepth]{subfig}
\usepackage{color}
\usepackage{color}

\begin{document}
\title{Synchronization of a periodic modulation of mirrors in an optomechanical system}
\author{Vahid Ameri}
\email{vahameri@gmail.com}
\affiliation{Department of Physics, Faculty of Science, University of Hormozgan, Bandar-Abbas, Iran}
\author{Mohammad Eghbali-Arani}
\affiliation{Department of Physics, University of Kashan, Kashan, Iran }
\author{Morteza Rafiee}
\affiliation{Department of Physics, Shahrood University of Technology, 3619995161 Shahrood, Iran}
\begin{abstract}
Proposing an optomechanical cavity modulated periodically, we study the modulation synchronization of mechanical modes of the mirrors. A periodic modulation is applied to one of the mirrors, where the second mirror has the capability of oscillation, without any modulation for that.  As a result, we find a phase-locking synchronization between the mechanical modes of the mirrors and enhancement of quantum synchronization by having the periodic modulation.  Using the fact that, periodic modulation can make the squeezed states, we show that there is a robust synchronization of periodic modulation between mirrors against enhancement of detuning between the mirrors. Also, our results show that having a periodic modulation leads to a stationary entanglement generation between the mirrors.
\end{abstract}
\maketitle
\section{Introduction}
Synchronization as a well-known classical phenomenon has been observed in a large variety of contexts, eg. physical systems, biology, chemical reactions, and etc.  \cite{pikovsky2003synchronization, meyr1997digital, fries2001modulation, selinger2004measuring, simon2012situ}. The destructive effect of noises on synchronization and the important role of them in the quantum regime made the quantum systems very attractive to be considered as a great system to trace the effect of noises on synchronization \cite{walter2014quantum,mari2013measures,li2016quantum}. Although finding a proper and global measure of quantum synchronization, and also the absence of a clear notion of phase space trajectories are still the main challenges in the studies on quantum synchronization \cite{mari2013measures}.
\par As mentioned before, noise statistic overwhelms synchronization in the quantum regime. So any scenarios of reducing the noises could be an interesting proposal of making a synchronization. S.Sonew and et all, using the squeezing hamiltonian instead of a harmonic derive, show that the squeezing enhances quantum synchronization \cite{sonar2018squeezing}. 
\par Periodic modulation has been reported as a technique which may increase the possibility of achievement of squeezed states in a wide variety of optomechanical systems \cite{purdy2013strong, clerk2008back, huang2009enhancement, ian2008cavity, jahne2009cavity}. By choosing appropriate periodic modulation, one can achieve better squeezing in compare to the other method of the creation of squeezed states \cite{farace2012enhancing, mari2009gently}.  
\par The effect of periodic modulation on quantum systems are widely studied \cite{goldman2014periodically, wang2006periodic, grifoni1998driven}. A. Frace and V. Giovannetti showed enhancement on quantum effects in optomechanical systems using periodic modulation \cite{farace2012enhancing}. Recently a synchronization enhancement via periodic modulation in optomechanical systems is also reported \cite{du2017synchronization}.
\par The aim of this work is to estimate the synchronization of the periodic oscillation of the mirrors, raised by the periodic modulation of one of them, in an optomechanical system. It is also interesting to study the correlations between the mirrors and finding a relation between correlations and synchronization.
\par In this paper, we propose an optomechanical system with two oscillating mirrors coupled to the same optical field in a cavity. One of the mirrors (M1) is considered to be modulated periodically (Fig.1). We investigate the generation of a squeezed state of the mirrors and also the synchronization and phase locking between their oscillation raised by the modulation.
\par In following, our time modulated optomechanical system is introduced and the Hamiltonian and the master equation of the system are discussed. Then, using QuTiP\cite{johansson2012qutip} to solve the master equation and also Langevin
equations, we derive the dynamics of the system and discuss the results.
\begin{figure}
\includegraphics[width=0.8 \columnwidth ]{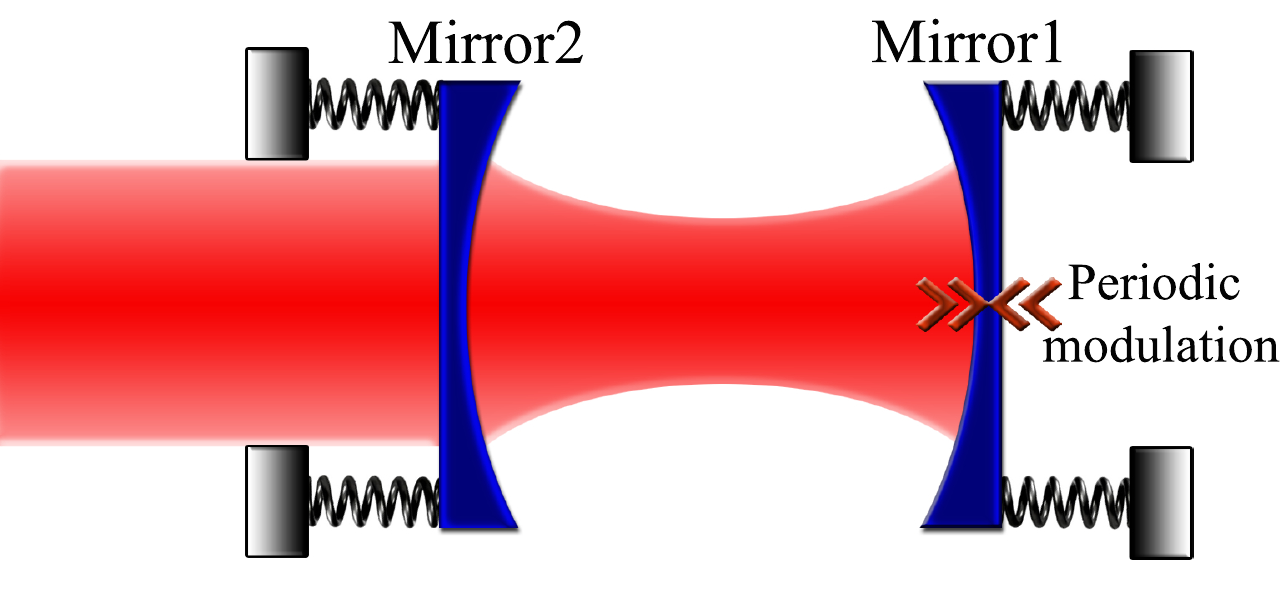}\\
\caption{(Color online) The system consists of a Fabry-Perot cavity driven by an external laser and two movable mirrors(M1 and M2), where the right mirror(M1) is modulated periodically.}\label{setup}
\end{figure}
\section{The system}
\par We proposed an optomechanical cavity with two movable mirrors at both ends where they are interacting with the optical field of the cavity of frequency $ \omega_c $. The mirrors are attached to springs of effective characteristic frequencies $ \omega_{M} $ and $ \omega_{M}+\Delta_M $ for mirror1(M1) and mirror2(M2) respectively, where $ \Delta_M $ is assumed to be the detuning of the frequencies of the mirrors. The cavity is driven by a laser of frequency $ \omega_L $ as it can be seen in Fig. \ref{setup}. The periodic modulation on M1 is expressed as a time-dependent function in the Hamiltonian of the system,
\begin{eqnarray}\label{h}
H = \Delta a^\dagger a+\frac{\omega_M}{2}p_1^2++\frac{\omega_M}{2}q_1^2[1+\epsilon \sin ^2(\Omega t)]  \\
+\frac{(\omega_M+\Delta_M)}{2}(p_2^2+q_2^2)- g a^\dagger a(q_1+q_2) +i E( a^\dagger - a) \nonumber
\end{eqnarray}
where $ \hbar =1 $, $ E $ determines the strength of the external drive, $ q_{1,2} $ and $ p_{1,2} $ are the dimensionless position and momentum of the mirrors satisfying the commutation relation $ [q_k,p_j]=i\delta_{jk} $ and $ g $ is the mirrors-field coupling constant. In addition, $ \Delta=\omega_c-\omega_L $ is the detuning, and $ a(a^\dagger) $ is the annihilation(creation) operator, of the cavity mode. Also, $ \epsilon $ determines the domain of the modulation and $ \Omega $ is the modulation frequency. 

\par The full dynamics of the system is described by a set of nonlinear Langevin
equations, including the effects of vacuum radiation noise and quantum Brownian
noise acting on the mirror. Dynamics of the cavity and mirror modes under the Hamiltonian (\ref{h}), are given by

\begin{equation}
\begin{aligned}
\dot{a}&=-(i \Delta + \kappa) a + i g (q_1+q_2) a + E+ \sqrt{2\kappa} a^{in},\\
\dot{q_1}&= \omega_M p_1, \\
\dot{p_1}&= -\omega_M(1+\epsilon \sin ^2(\Omega t))q_1 +g a^\dagger a - \gamma_{m1} p_1 + \xi_1,\\
\dot{q_2}&= (\omega_M + \Delta_M) p_2, \\
\dot{p_2}&= -(\omega_M + \Delta_M) q_2 +g a^\dagger a - \gamma_{m2} p_2 + \xi_2,\\
\label{Langevin_1}
\end{aligned}
\end{equation}

where $a^{in} $ is the radiation vacuum input noise with autocorrelation function $\langle a^{in}(t) a^{in \dagger} (t') \rangle=\delta(t-t')$ and $\xi_i(t)$ is the Brownian noise operator of the mirror which is for a good quality mirror with $\omega_M \gg \gamma_m$ has autocorrelation function satisfies the relation \cite{giovannetti2001phase}
\begin{equation}
\langle \{\xi(t) , \xi(t') \} \rangle \approx 2 \gamma_m \coth(\frac{\hbar \omega_M}{2 K_B T}) \delta(t-t'),
\end{equation}
where in the above equation, $\langle . \rangle$ is the anticomutator, $K_B$ is the Boltzmann constant and $T$ is the system temperature. 
Meanwhile, each operator can be expressed as sum of stationary value plus an additional fluctuation operator, $a=\alpha_s+\delta a$, $a^\dagger=\alpha_s+\delta a^\dagger$, $q_i=Q_i+\delta q_i$ and $p_i=P_i+\delta p_i$, while $Q_i (P_i), i=1,2$ is steady values of dimensionless position (momentum). With definition $x=\frac{a+a^\dagger}{\sqrt{2}}$, $y=\frac{a-a^\dagger}{i \sqrt{2}}$, $x^{in}=\frac{a^{in}+a^{in \dagger}}{\sqrt{2}}$, $y^{in}=\frac{a^{in}-a^{in \dagger}}{i \sqrt{2}}$ are the quadratures of the field and input noises, substitution of these equations in the regime with large coherent amplitudes for mechanical and optical field leads to the exact quantum linear Langevin equations for the fluctuations as
\begin{equation}
\begin{aligned}
\delta\dot{ x}&= \Delta \delta y - \kappa \delta x + \sqrt{2\kappa} \delta x^{in},\\
\delta\dot{ y}&= -\Delta \delta x - \kappa \delta y + G (\delta q_1 + \delta q_2) + \sqrt{2\kappa} \delta y^{in},\\
\delta\dot{ q_1}&= \omega_M \delta p_1, \\
\delta\dot{ p_1}&= -\omega_M(1+\epsilon \sin ^2(\Omega t)) \delta q_1 + G \delta x - \gamma_{m1} \delta p_1 + \xi_1,\\
\delta\dot{ q_2}&= (\omega_M + \Delta_M) \delta p_2, \\
\delta\dot{ p_2}&= -(\omega_M + \Delta_M) \delta q_2 + G \delta x  - \gamma_{m2} \delta p_2 + \xi_2,\\
\label{Langevin_2}
\end{aligned}
\end{equation}

where $ G=\sqrt{2} g \alpha_s $. It could be possible to rewrite above equations of motions in the compact matrix form

\begin{equation}\label{eq_motion}
\delta \dot{U(t)}=A .\delta U(t) + \delta N^{in}(t).
\end{equation}

where $\delta U(t)=(\delta x(t), \delta y(t), \delta q_1(t), \delta p_1(t), \delta q_2(t), \delta p_2(t))^T$ is the system operator vector and $\delta N^{in}(t)=( \delta x^{in},  \delta Y^{in},  0, \xi_1, 0, \xi_2)^T$  is the vector of noises and the drift matrix A is given by
\begin{widetext}
\begin{equation}\label{10}
A =\left(\begin{array}{*{20}c}
{{-\kappa}} & {{\Delta}} & {{0}} & {{0}} & {{0}} & {{0}}  \\
{{\Delta}} & {{-\kappa}} & {{G}} & {{0}}  & {{G}} & {{0}}\\
{{0}} & {{0}} &  {{0}} & {{\omega_m}} & {{0}} & {{0}} \\
{{G}} & {{0}} & {{-\omega_M(1+\epsilon \sin ^2(\Omega t))}} & {{-\gamma_{m1}}} & {{0}} & {{0}}  \\
{{0}} & {{0}} &  {{0}} & {{0}} & {{0}} & {{\omega_M+\Delta_M}} \\
{{G}} & {{0}} & {{0}} & {{0}} & {{-(\omega_M+\Delta_M)}} & {{-\gamma_{m2}}}  \\
\end{array}\right).
\end{equation}
\end{widetext}
Complete description of the system is given in terms of the second
statistical moments, which can be arranged in the covariance matrix $\sigma$ of entries
$\sigma_{ij}(t):=\langle \{ U_i(t) U_j(t) \} \rangle /2 -\langle U_i(t) \rangle \langle U_j(t) \rangle$. The equation of motion for the covariance matrix ($\sigma$) is
\begin{equation}
\dot{\sigma}= A \sigma + \sigma A^T + D,
\end{equation}
where the diffusion matrix $D$ reads $D=diag \{\kappa (2n_{ph}+1), \kappa (2n_{ph}+1), 0, \gamma_{m1} (2n_{m1}+1), 0, \gamma_{m2} (2n_{m2}+1) \}$ while $n_m=(e^{\hbar \omega_m/k_B T}-1)^{-1}$ ($n_{ph}=(e^{\hbar \Delta/k_B T}-1)^{-1}$) is the mean phonon number of the mirror (field). These calculations have been done using the linearization technique while it is possible to obtain the dynamics of the system by solving the master equation of the system directly. In the following, we introduce the master equation governing the dynamics of our system and it would be worth to check the amounts that the linearization affects our results by comparing them from both approaches.  
\par Also, one can use the master equation approach to derive the dynamics of the system,
\begin{eqnarray}
\frac{d\rho}{dt}=-i\left[ \rho , H \right] +\sum_{i=1}^2\gamma_i \left( 2b_i\rho b_i^\dagger- b_i^\dagger b_i\rho - \rho  b_i^\dagger b_i\right) \\ \nonumber
+\kappa \left( 2a\rho a^\dagger-a_i^\dagger a\rho - \rho  a^\dagger a\right).
\end{eqnarray}
Where $ b_i(1,2) $ is the annihilation operator corresponds to the mechanical mode of the mirrors and system is intended to be at zero temperature. Mirrors are also considered to have the same damping rate $ \gamma_i (1,2)$ while $ \kappa $ is the decay rate of the cavity mode.
\section{Results}
We followed both, master equation and Langevin equations, approaches. 
 Setting $\omega_M=1$, other used normalized parameters to $\omega_M$ are as  $\Delta = 1 $, $ \kappa = 0.1 $, $ \gamma_1=\gamma_2 = 0.001 $, $ g= 0.05 $, $ \Omega=0.5 $, $ \epsilon=0.5 $ and $ E=2.1 $. The detuning $ \Delta_M $ is left as variable to study the robustness of synchronization.
\subsection{Squeezing}
\begin{figure}
\includegraphics[width=0.95 \columnwidth ]{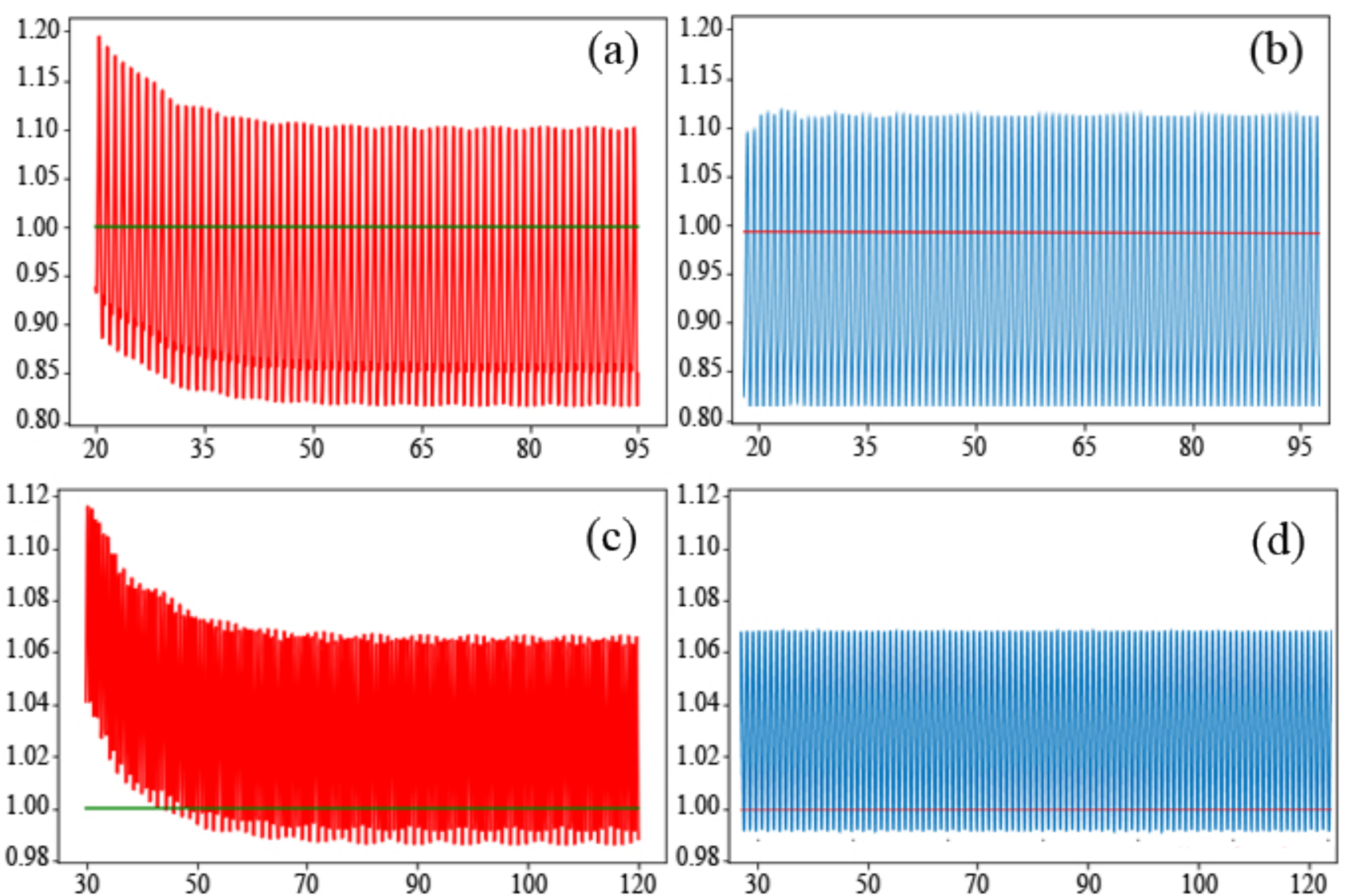}\\
\caption{(Color online) Fluctuations after a sufficiently long time when the system is almost stable as a function of $t/\tau (\tau=\frac{2 \pi}{\omega}) $.  $ <\Delta q_1^2>/<\Delta q_0^2> $ (a) from master equation approach (b) obtained by Langevin equations and covariance matrix, $ <\Delta q_2^2>/<\Delta q_0^2> $, (c) from master equation approach (d) obtained by Langevin equations and covariance matrix, where $ q_0 $ is the zero point level of position quadrature.}\label{sq}
\end{figure}
In a quantum regime, fluctuations play a more important role. It is expected to overcome the complications associated with added noises originating from quantum fluctuations by squeezing and enhance quantum synchronization \cite{sonar2018squeezing}. Periodic modulation is introduced as a practical approach to preparing squeezed states. Since it is an oscillating behaviour, it would be interesting to study the synchronization of that.  Indeed the first sign of having periodic modulation synchronization, would be the squeezing of M2. Indeed, the first sign of having periodic modulation synchronization would be the squeezing of the states of M2.
\par The ground-state fluctuations minimize the uncertainty relation,
\begin{equation}
<\Delta X_1^2><\Delta X_2^2>\geqslant \frac{1}{4}\mid<[X_1,X_2]> \mid^2.
\end{equation}
It is possible to squeeze one of the quadratures to have the fluctuations below the zero-point level at the expense of increasing the fluctuation of the other quadrature. 
\par Using the periodic modulation $\omega_M q_1^2 \epsilon \sin ^2(\Omega t) $, in Hamiltonian Eq.\ref{h}, the states of M1 are squeezed.Although only M1 is modulated, in case of zero-detuning between the mirrors $ \Delta_M=0 $, the states of M2  are also squeezed (Fig.\ref{sq}). One can consider this squeezing of the states of M2 as a result of periodic modulation synchronization of the mirrors.
As can be seen from these results, for our system, there is no significant difference between both the numerical approaches. Since all the results have a great coincidence, we will demonstrate just the results obtained by the master equation approach using QuTip.
\subsection{Synchronization}
\begin{figure}
\includegraphics[width=0.8 \columnwidth ]{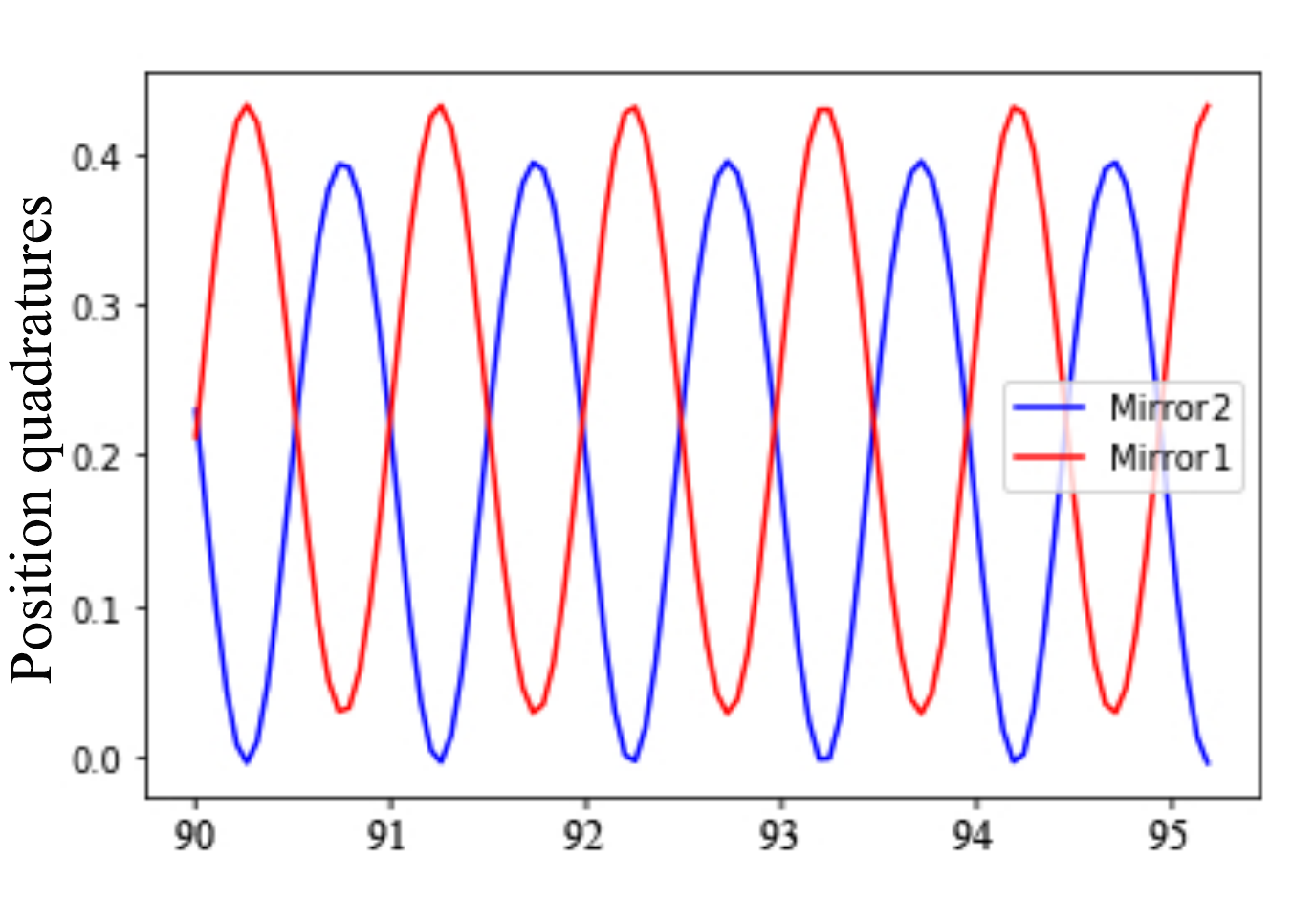}\\
\caption{(Color online) Oscillations of position quadratures of the mirrors after a sufficiently long time as a function of $t/\tau (\tau=\frac{2 \pi}{\omega}) $.}\label{osc0}
\end{figure}

\begin{figure}
\includegraphics[width=0.85 \columnwidth ]{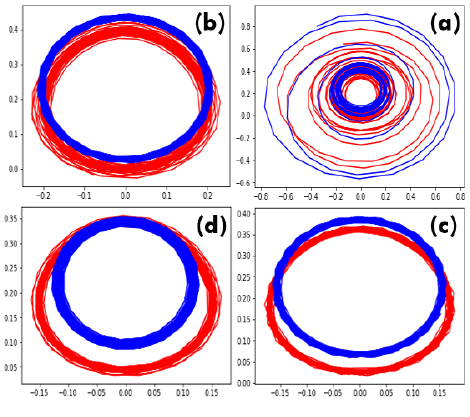}\\
\caption{(Color online) The evolution of position and momentum quadratures of the mirrors for a) $ \Delta_M=0 $ at the beginging, and b) $ \Delta_M=0 $, c) $ \Delta_M=0.05 $, d) $ \Delta_M=0.1 $  after a sufficiently long time.}\label{cyc0}
\end{figure}

Before using any measure to determine the synchronization and phase locking between the oscillation of the mirrors, supported and enhanced by periodic modulation of M1, we can take a look at Fig.\ref{osc0} where the final oscillations of the position quadratures $ q_1 $ and $ q_2 $ are plotted. It is easy to find a phase locking between the oscillations of the mirrors. Also, it is interesting to have the evolution of the position and momentum quadratures of the mirrors. In Fig.\ref{cyc0}, one can see that the evolution tends to a periodic orbit for both of mirrors. In case of zero-detuning ($ \Delta_M=0 $), the orbits have almost same dimension while considering the non zero-detuning regime the second mirror orbit get smaller than the first mirror by increasing the detuning $ \Delta_M $.
\par Mari et al. \cite{mari2013measures} introduced the following measure for synchronization that one can gauge the synchronization level of two subsystem using their position and momentum quadratures,
\begin{equation}
S(t)=<q_-^2(t)+p_-^2>^{-1},
\end{equation}
where $ q_-(t)=[q_1-q_2]/\sqrt{2} $ and $ p_-(t)=[p_1-p_2]/\sqrt{2} $. The synchronization measure $ S(t) $ has a maximal value 1.0 correspond to a complete synchronization. This limit is applied on $ S(t) $ by Heisenberg's uncertainty principle.
\par Lei Du et al. enhanced a quantum synchronization in an optomechanical system using a proper periodic modulation \cite{du2017synchronization}. They achieved the synchronization measure $ S(t) $ up to 0.9. We find the same value of  $ S(t) $ for the synchronization of the periodic modulation in our optomechanical system.
\begin{figure}
\includegraphics[width=0.9 \columnwidth ]{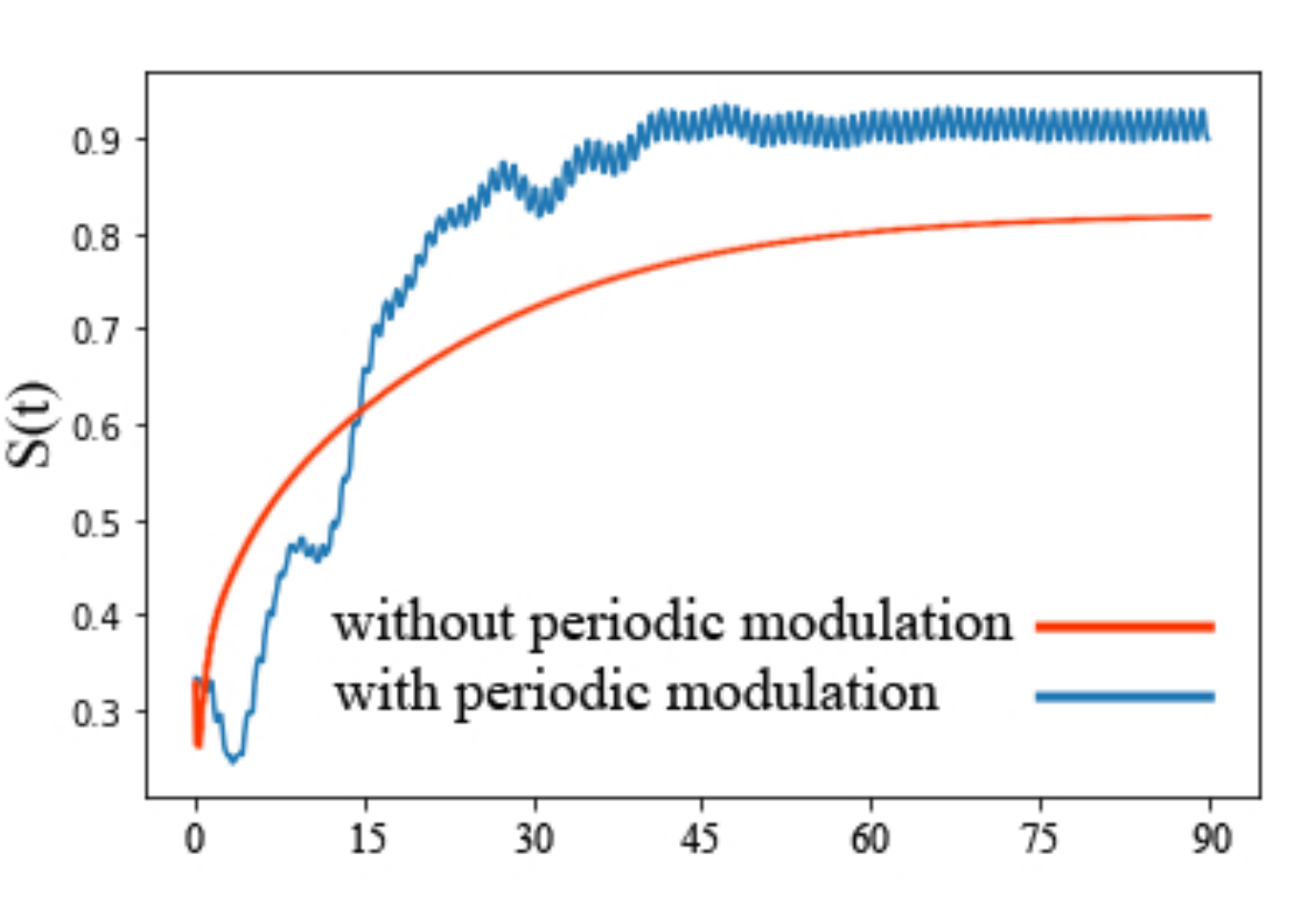}\\
\caption{(Color online) Quantum synchronization measure $ S(t) $ as a function of $t/\tau (\tau=\frac{2 \pi}{\omega})  $ for periodically modulated system and $\times 80 \; t/\tau  $ for the system without periodic modulation.}\label{sc}
\end{figure}
\subsection{Mutual information and entanglement }
The study of synchronization in a quantum system composed of two subsystems A and B where they are correlated attracts a lot of interests  \cite{manzano2013synchronization, giorgi2012quantum, giorgi2013spontaneous}. For instance, entanglement between synchronized subsystems has been studied in a variety of systems \cite{giovannetti2002positioning, lee2014entanglement, zhirov2009quantum}. 
\par
\begin{figure}
\includegraphics[width=0.8 \columnwidth ]{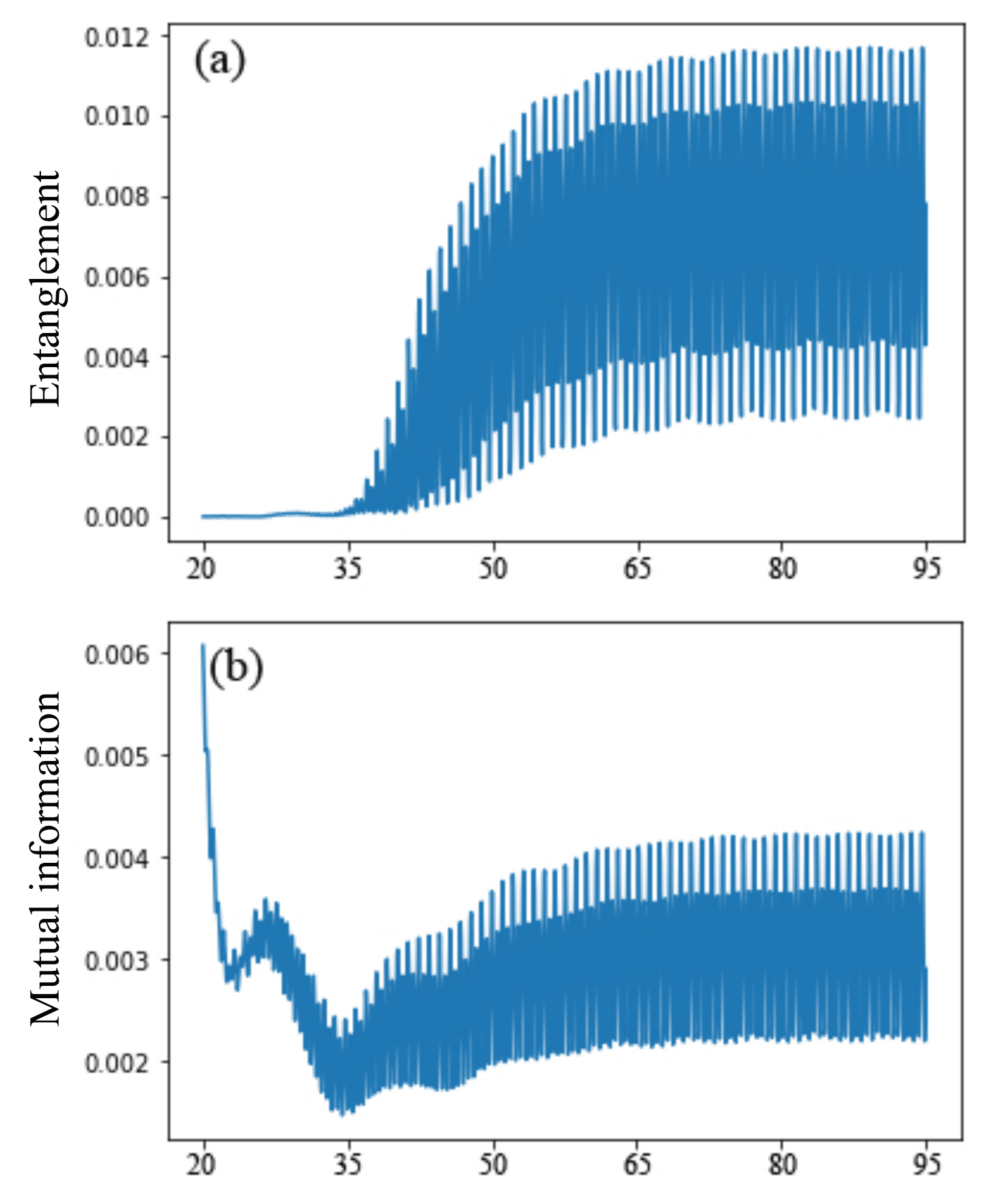}\\
\caption{(Color online) a) Entanglement and b) mutual information as a function of $t/\tau (\tau=\frac{2 \pi}{\omega}) $.}\label{ent}
\end{figure}

 Among the variety of entanglement measures, we used  logarithmic negativity is defined as,
\begin{equation}
E_N(\rho)= \log_2(|| \rho^{\Gamma_A}||),
\end{equation}
where $ \Gamma_A $ is the partial transpose operation with respect to subsystem $ A $ and $ ||.|| $ denotes the trace norm. In Fig.\ref{ent}(a) the logarithmic negativity of the mirrors are plotted as a function of time. As a result of periodic modulation, the mirrors get entangled. Very recently Roulet, et al. studied the relation between quantum synchronization and the generation of entanglement \cite{roulet2018quantum}. Comparing Fig.\ref{ent}(a) and Fig.\ref{sc}, one can conclude that the entanglement generation starts as the mirrors are getting synchronized.   Although the relation between entanglement and synchronization strongly depends on the details of the system, the mutual information can signal about the presence of a quantum synchronization more generally.
Recently, Ameri et al. showed that the synchronized subsystems have large mutual information. So, it can be used as a signal of quantum synchronization \cite{ameri2015mutual}. The quantum mutual information is defined as,
\begin{equation}
I = S(\rho_A)+S(\rho_B)+S(\rho),
\end{equation} 
where the quantum state of the system is traced partially to derive the reduced density matrices $\rho_A={\rm Tr}_B (\rho)$ and $\rho_B={\rm Tr}_A (\rho)$ and $S(\rho)=-{\rm Tr}[\rho \log(\rho)]$ is the Von Neumann entropy. 
Fig.\ref{ent}(b) shows the mutual information of mirrors as a function of time.  As can be seen from the Fig.\ref{sc} and Fig.\ref{ent}, the strong correlation between the mutual information, entanglement and synchronization approved again. 
\section{Conclusion}
In conclusion, we have demonstrated the possibility of synchronizing a periodic modulation of one mirror to another one in an optomechanical system. As a result, it has been shown that the state of the second mirror is also squeezed where it is one of the consequences of periodically modulated mirrors and the evolution of position and momentum quadratures for both mirrors follow the same periodic orbits originating from the periodic modulation of the first mirror. Finally, the quantum synchronization measure $ S(t) $ indicates that the mirrors are almost completely synchronized.
\par Another interesting result of this work is an entanglement generation between the mirrors due to the periodic modulation. Looking at Fig.\ref{ent} and Fig.\ref{sc}, one can conclude that the entanglement generation and the synchronization are almost started at the same time. So, as expected and reported previously, the correlations can signal us about the presence of synchronization in the systems.


\begin{acknowledgments}
This work was supported by University of Kashan and University of Hormozgan through the Project No. 96/8423.
\end{acknowledgments}
\bibliography{sci}

\begin{thebibliography}{30}%
\makeatletter
\providecommand \@ifxundefined [1]{%
 \@ifx{#1\undefined}
}%
\providecommand \@ifnum [1]{%
 \ifnum #1\expandafter \@firstoftwo
 \else \expandafter \@secondoftwo
 \fi
}%
\providecommand \@ifx [1]{%
 \ifx #1\expandafter \@firstoftwo
 \else \expandafter \@secondoftwo
 \fi
}%
\providecommand \natexlab [1]{#1}%
\providecommand \enquote  [1]{``#1''}%
\providecommand \bibnamefont  [1]{#1}%
\providecommand \bibfnamefont [1]{#1}%
\providecommand \citenamefont [1]{#1}%
\providecommand \href@noop [0]{\@secondoftwo}%
\providecommand \href [0]{\begingroup \@sanitize@url \@href}%
\providecommand \@href[1]{\@@startlink{#1}\@@href}%
\providecommand \@@href[1]{\endgroup#1\@@endlink}%
\providecommand \@sanitize@url [0]{\catcode `\\12\catcode `\$12\catcode
  `\&12\catcode `\#12\catcode `\^12\catcode `\_12\catcode `\%12\relax}%
\providecommand \@@startlink[1]{}%
\providecommand \@@endlink[0]{}%
\providecommand \url  [0]{\begingroup\@sanitize@url \@url }%
\providecommand \@url [1]{\endgroup\@href {#1}{\urlprefix }}%
\providecommand \urlprefix  [0]{URL }%
\providecommand \Eprint [0]{\href }%
\providecommand \doibase [0]{http://dx.doi.org/}%
\providecommand \selectlanguage [0]{\@gobble}%
\providecommand \bibinfo  [0]{\@secondoftwo}%
\providecommand \bibfield  [0]{\@secondoftwo}%
\providecommand \translation [1]{[#1]}%
\providecommand \BibitemOpen [0]{}%
\providecommand \bibitemStop [0]{}%
\providecommand \bibitemNoStop [0]{.\EOS\space}%
\providecommand \EOS [0]{\spacefactor3000\relax}%
\providecommand \BibitemShut  [1]{\csname bibitem#1\endcsname}%
\let\auto@bib@innerbib\@empty
\bibitem [{\citenamefont {Pikovsky}\ \emph {et~al.}(2003)\citenamefont
  {Pikovsky}, \citenamefont {Rosenblum}, \citenamefont {Kurths},\ and\
  \citenamefont {Kurths}}]{pikovsky2003synchronization}%
  \BibitemOpen
  \bibfield  {author} {\bibinfo {author} {\bibfnamefont {A.}~\bibnamefont
  {Pikovsky}}, \bibinfo {author} {\bibfnamefont {M.}~\bibnamefont {Rosenblum}},
  \bibinfo {author} {\bibfnamefont {J.}~\bibnamefont {Kurths}}, \ and\ \bibinfo
  {author} {\bibfnamefont {J.}~\bibnamefont {Kurths}},\ }\href@noop {} {\emph
  {\bibinfo {title} {Synchronization: a universal concept in nonlinear
  sciences}}},\ Vol.~\bibinfo {volume} {12}\ (\bibinfo  {publisher} {Cambridge
  university press},\ \bibinfo {year} {2003})\BibitemShut {NoStop}%
\bibitem [{\citenamefont {Meyr}\ \emph {et~al.}(1997)\citenamefont {Meyr},
  \citenamefont {Moeneclaey},\ and\ \citenamefont {Fechtel}}]{meyr1997digital}%
  \BibitemOpen
  \bibfield  {author} {\bibinfo {author} {\bibfnamefont {H.}~\bibnamefont
  {Meyr}}, \bibinfo {author} {\bibfnamefont {M.}~\bibnamefont {Moeneclaey}}, \
  and\ \bibinfo {author} {\bibfnamefont {S.}~\bibnamefont {Fechtel}},\
  }\href@noop {} {\emph {\bibinfo {title} {Digital communication receivers:
  synchronization, channel estimation, and signal processing}}}\ (\bibinfo
  {publisher} {John Wiley \& Sons, Inc.},\ \bibinfo {year} {1997})\BibitemShut
  {NoStop}%
\bibitem [{\citenamefont {Fries}\ \emph {et~al.}(2001)\citenamefont {Fries},
  \citenamefont {Reynolds}, \citenamefont {Rorie},\ and\ \citenamefont
  {Desimone}}]{fries2001modulation}%
  \BibitemOpen
  \bibfield  {author} {\bibinfo {author} {\bibfnamefont {P.}~\bibnamefont
  {Fries}}, \bibinfo {author} {\bibfnamefont {J.~H.}\ \bibnamefont {Reynolds}},
  \bibinfo {author} {\bibfnamefont {A.~E.}\ \bibnamefont {Rorie}}, \ and\
  \bibinfo {author} {\bibfnamefont {R.}~\bibnamefont {Desimone}},\ }\href@noop
  {} {\bibfield  {journal} {\bibinfo  {journal} {Science}\ }\textbf {\bibinfo
  {volume} {291}},\ \bibinfo {pages} {1560} (\bibinfo {year}
  {2001})}\BibitemShut {NoStop}%
\bibitem [{\citenamefont {Selinger}\ \emph {et~al.}(2004)\citenamefont
  {Selinger}, \citenamefont {Pancrazio},\ and\ \citenamefont
  {Gross}}]{selinger2004measuring}%
  \BibitemOpen
  \bibfield  {author} {\bibinfo {author} {\bibfnamefont {J.~V.}\ \bibnamefont
  {Selinger}}, \bibinfo {author} {\bibfnamefont {J.~J.}\ \bibnamefont
  {Pancrazio}}, \ and\ \bibinfo {author} {\bibfnamefont {G.~W.}\ \bibnamefont
  {Gross}},\ }\href@noop {} {\bibfield  {journal} {\bibinfo  {journal}
  {Biosensors and Bioelectronics}\ }\textbf {\bibinfo {volume} {19}},\ \bibinfo
  {pages} {675} (\bibinfo {year} {2004})}\BibitemShut {NoStop}%
\bibitem [{\citenamefont {Simon}\ \emph {et~al.}(2012)\citenamefont {Simon},
  \citenamefont {Merz}, \citenamefont {Dubuis}, \citenamefont {Lieb},\ and\
  \citenamefont {Hungerbuhler}}]{simon2012situ}%
  \BibitemOpen
  \bibfield  {author} {\bibinfo {author} {\bibfnamefont {L.}~\bibnamefont
  {Simon}}, \bibinfo {author} {\bibfnamefont {T.}~\bibnamefont {Merz}},
  \bibinfo {author} {\bibfnamefont {S.}~\bibnamefont {Dubuis}}, \bibinfo
  {author} {\bibfnamefont {A.}~\bibnamefont {Lieb}}, \ and\ \bibinfo {author}
  {\bibfnamefont {K.}~\bibnamefont {Hungerbuhler}},\ }\href@noop {} {\bibfield
  {journal} {\bibinfo  {journal} {Chemical Engineering Research and Design}\
  }\textbf {\bibinfo {volume} {90}},\ \bibinfo {pages} {1847} (\bibinfo {year}
  {2012})}\BibitemShut {NoStop}%
\bibitem [{\citenamefont {Walter}\ \emph {et~al.}(2014)\citenamefont {Walter},
  \citenamefont {Nunnenkamp},\ and\ \citenamefont
  {Bruder}}]{walter2014quantum}%
  \BibitemOpen
  \bibfield  {author} {\bibinfo {author} {\bibfnamefont {S.}~\bibnamefont
  {Walter}}, \bibinfo {author} {\bibfnamefont {A.}~\bibnamefont {Nunnenkamp}},
  \ and\ \bibinfo {author} {\bibfnamefont {C.}~\bibnamefont {Bruder}},\
  }\href@noop {} {\bibfield  {journal} {\bibinfo  {journal} {Physical review
  letters}\ }\textbf {\bibinfo {volume} {112}},\ \bibinfo {pages} {094102}
  (\bibinfo {year} {2014})}\BibitemShut {NoStop}%
\bibitem [{\citenamefont {Mari}\ \emph {et~al.}(2013)\citenamefont {Mari},
  \citenamefont {Farace}, \citenamefont {Didier}, \citenamefont {Giovannetti},\
  and\ \citenamefont {Fazio}}]{mari2013measures}%
  \BibitemOpen
  \bibfield  {author} {\bibinfo {author} {\bibfnamefont {A.}~\bibnamefont
  {Mari}}, \bibinfo {author} {\bibfnamefont {A.}~\bibnamefont {Farace}},
  \bibinfo {author} {\bibfnamefont {N.}~\bibnamefont {Didier}}, \bibinfo
  {author} {\bibfnamefont {V.}~\bibnamefont {Giovannetti}}, \ and\ \bibinfo
  {author} {\bibfnamefont {R.}~\bibnamefont {Fazio}},\ }\href@noop {}
  {\bibfield  {journal} {\bibinfo  {journal} {Physical review letters}\
  }\textbf {\bibinfo {volume} {111}},\ \bibinfo {pages} {103605} (\bibinfo
  {year} {2013})}\BibitemShut {NoStop}%
\bibitem [{\citenamefont {Li}\ \emph {et~al.}(2016)\citenamefont {Li},
  \citenamefont {Li},\ and\ \citenamefont {Song}}]{li2016quantum}%
  \BibitemOpen
  \bibfield  {author} {\bibinfo {author} {\bibfnamefont {W.}~\bibnamefont
  {Li}}, \bibinfo {author} {\bibfnamefont {C.}~\bibnamefont {Li}}, \ and\
  \bibinfo {author} {\bibfnamefont {H.}~\bibnamefont {Song}},\ }\href@noop {}
  {\bibfield  {journal} {\bibinfo  {journal} {Physical Review E}\ }\textbf
  {\bibinfo {volume} {93}},\ \bibinfo {pages} {062221} (\bibinfo {year}
  {2016})}\BibitemShut {NoStop}%
\bibitem [{\citenamefont {Sonar}\ \emph {et~al.}(2018)\citenamefont {Sonar},
  \citenamefont {Hajdu{\v{s}}ek}, \citenamefont {Mukherjee}, \citenamefont
  {Fazio}, \citenamefont {Vedral}, \citenamefont {Vinjanampathy},\ and\
  \citenamefont {Kwek}}]{sonar2018squeezing}%
  \BibitemOpen
  \bibfield  {author} {\bibinfo {author} {\bibfnamefont {S.}~\bibnamefont
  {Sonar}}, \bibinfo {author} {\bibfnamefont {M.}~\bibnamefont
  {Hajdu{\v{s}}ek}}, \bibinfo {author} {\bibfnamefont {M.}~\bibnamefont
  {Mukherjee}}, \bibinfo {author} {\bibfnamefont {R.}~\bibnamefont {Fazio}},
  \bibinfo {author} {\bibfnamefont {V.}~\bibnamefont {Vedral}}, \bibinfo
  {author} {\bibfnamefont {S.}~\bibnamefont {Vinjanampathy}}, \ and\ \bibinfo
  {author} {\bibfnamefont {L.-C.}\ \bibnamefont {Kwek}},\ }\href@noop {}
  {\bibfield  {journal} {\bibinfo  {journal} {Physical review letters}\
  }\textbf {\bibinfo {volume} {120}},\ \bibinfo {pages} {163601} (\bibinfo
  {year} {2018})}\BibitemShut {NoStop}%
\bibitem [{\citenamefont {Purdy}\ \emph {et~al.}(2013)\citenamefont {Purdy},
  \citenamefont {Yu}, \citenamefont {Peterson}, \citenamefont {Kampel},\ and\
  \citenamefont {Regal}}]{purdy2013strong}%
  \BibitemOpen
  \bibfield  {author} {\bibinfo {author} {\bibfnamefont {T.}~\bibnamefont
  {Purdy}}, \bibinfo {author} {\bibfnamefont {P.-L.}\ \bibnamefont {Yu}},
  \bibinfo {author} {\bibfnamefont {R.}~\bibnamefont {Peterson}}, \bibinfo
  {author} {\bibfnamefont {N.}~\bibnamefont {Kampel}}, \ and\ \bibinfo {author}
  {\bibfnamefont {C.}~\bibnamefont {Regal}},\ }\href@noop {} {\bibfield
  {journal} {\bibinfo  {journal} {Physical Review X}\ }\textbf {\bibinfo
  {volume} {3}},\ \bibinfo {pages} {031012} (\bibinfo {year}
  {2013})}\BibitemShut {NoStop}%
\bibitem [{\citenamefont {Clerk}\ \emph {et~al.}(2008)\citenamefont {Clerk},
  \citenamefont {Marquardt},\ and\ \citenamefont {Jacobs}}]{clerk2008back}%
  \BibitemOpen
  \bibfield  {author} {\bibinfo {author} {\bibfnamefont {A.}~\bibnamefont
  {Clerk}}, \bibinfo {author} {\bibfnamefont {F.}~\bibnamefont {Marquardt}}, \
  and\ \bibinfo {author} {\bibfnamefont {K.}~\bibnamefont {Jacobs}},\
  }\href@noop {} {\bibfield  {journal} {\bibinfo  {journal} {New Journal of
  Physics}\ }\textbf {\bibinfo {volume} {10}},\ \bibinfo {pages} {095010}
  (\bibinfo {year} {2008})}\BibitemShut {NoStop}%
\bibitem [{\citenamefont {Huang}\ and\ \citenamefont
  {Agarwal}(2009)}]{huang2009enhancement}%
  \BibitemOpen
  \bibfield  {author} {\bibinfo {author} {\bibfnamefont {S.}~\bibnamefont
  {Huang}}\ and\ \bibinfo {author} {\bibfnamefont {G.}~\bibnamefont
  {Agarwal}},\ }\href@noop {} {\bibfield  {journal} {\bibinfo  {journal}
  {Physical Review A}\ }\textbf {\bibinfo {volume} {79}},\ \bibinfo {pages}
  {013821} (\bibinfo {year} {2009})}\BibitemShut {NoStop}%
\bibitem [{\citenamefont {Ian}\ \emph {et~al.}(2008)\citenamefont {Ian},
  \citenamefont {Gong}, \citenamefont {Liu}, \citenamefont {Sun},\ and\
  \citenamefont {Nori}}]{ian2008cavity}%
  \BibitemOpen
  \bibfield  {author} {\bibinfo {author} {\bibfnamefont {H.}~\bibnamefont
  {Ian}}, \bibinfo {author} {\bibfnamefont {Z.}~\bibnamefont {Gong}}, \bibinfo
  {author} {\bibfnamefont {Y.-x.}\ \bibnamefont {Liu}}, \bibinfo {author}
  {\bibfnamefont {C.}~\bibnamefont {Sun}}, \ and\ \bibinfo {author}
  {\bibfnamefont {F.}~\bibnamefont {Nori}},\ }\href@noop {} {\bibfield
  {journal} {\bibinfo  {journal} {Physical Review A}\ }\textbf {\bibinfo
  {volume} {78}},\ \bibinfo {pages} {013824} (\bibinfo {year}
  {2008})}\BibitemShut {NoStop}%
\bibitem [{\citenamefont {J{\"a}hne}\ \emph {et~al.}(2009)\citenamefont
  {J{\"a}hne}, \citenamefont {Genes}, \citenamefont {Hammerer}, \citenamefont
  {Wallquist}, \citenamefont {Polzik},\ and\ \citenamefont
  {Zoller}}]{jahne2009cavity}%
  \BibitemOpen
  \bibfield  {author} {\bibinfo {author} {\bibfnamefont {K.}~\bibnamefont
  {J{\"a}hne}}, \bibinfo {author} {\bibfnamefont {C.}~\bibnamefont {Genes}},
  \bibinfo {author} {\bibfnamefont {K.}~\bibnamefont {Hammerer}}, \bibinfo
  {author} {\bibfnamefont {M.}~\bibnamefont {Wallquist}}, \bibinfo {author}
  {\bibfnamefont {E.~S.}\ \bibnamefont {Polzik}}, \ and\ \bibinfo {author}
  {\bibfnamefont {P.}~\bibnamefont {Zoller}},\ }\href@noop {} {\bibfield
  {journal} {\bibinfo  {journal} {Physical Review A}\ }\textbf {\bibinfo
  {volume} {79}},\ \bibinfo {pages} {063819} (\bibinfo {year}
  {2009})}\BibitemShut {NoStop}%
\bibitem [{\citenamefont {Farace}\ and\ \citenamefont
  {Giovannetti}(2012)}]{farace2012enhancing}%
  \BibitemOpen
  \bibfield  {author} {\bibinfo {author} {\bibfnamefont {A.}~\bibnamefont
  {Farace}}\ and\ \bibinfo {author} {\bibfnamefont {V.}~\bibnamefont
  {Giovannetti}},\ }\href@noop {} {\bibfield  {journal} {\bibinfo  {journal}
  {Physical Review A}\ }\textbf {\bibinfo {volume} {86}},\ \bibinfo {pages}
  {013820} (\bibinfo {year} {2012})}\BibitemShut {NoStop}%
\bibitem [{\citenamefont {Mari}\ and\ \citenamefont
  {Eisert}(2009)}]{mari2009gently}%
  \BibitemOpen
  \bibfield  {author} {\bibinfo {author} {\bibfnamefont {A.}~\bibnamefont
  {Mari}}\ and\ \bibinfo {author} {\bibfnamefont {J.}~\bibnamefont {Eisert}},\
  }\href@noop {} {\bibfield  {journal} {\bibinfo  {journal} {Physical Review
  Letters}\ }\textbf {\bibinfo {volume} {103}},\ \bibinfo {pages} {213603}
  (\bibinfo {year} {2009})}\BibitemShut {NoStop}%
\bibitem [{\citenamefont {Goldman}\ and\ \citenamefont
  {Dalibard}(2014)}]{goldman2014periodically}%
  \BibitemOpen
  \bibfield  {author} {\bibinfo {author} {\bibfnamefont {N.}~\bibnamefont
  {Goldman}}\ and\ \bibinfo {author} {\bibfnamefont {J.}~\bibnamefont
  {Dalibard}},\ }\href@noop {} {\bibfield  {journal} {\bibinfo  {journal}
  {Physical review X}\ }\textbf {\bibinfo {volume} {4}},\ \bibinfo {pages}
  {031027} (\bibinfo {year} {2014})}\BibitemShut {NoStop}%
\bibitem [{\citenamefont {Wang}\ \emph {et~al.}(2006)\citenamefont {Wang},
  \citenamefont {Fu},\ and\ \citenamefont {Liu}}]{wang2006periodic}%
  \BibitemOpen
  \bibfield  {author} {\bibinfo {author} {\bibfnamefont {G.-F.}\ \bibnamefont
  {Wang}}, \bibinfo {author} {\bibfnamefont {L.-B.}\ \bibnamefont {Fu}}, \ and\
  \bibinfo {author} {\bibfnamefont {J.}~\bibnamefont {Liu}},\ }\href@noop {}
  {\bibfield  {journal} {\bibinfo  {journal} {Physical Review A}\ }\textbf
  {\bibinfo {volume} {73}},\ \bibinfo {pages} {013619} (\bibinfo {year}
  {2006})}\BibitemShut {NoStop}%
\bibitem [{\citenamefont {Grifoni}\ and\ \citenamefont
  {H{\"a}nggi}(1998)}]{grifoni1998driven}%
  \BibitemOpen
  \bibfield  {author} {\bibinfo {author} {\bibfnamefont {M.}~\bibnamefont
  {Grifoni}}\ and\ \bibinfo {author} {\bibfnamefont {P.}~\bibnamefont
  {H{\"a}nggi}},\ }\href@noop {} {\bibfield  {journal} {\bibinfo  {journal}
  {Physics Reports}\ }\textbf {\bibinfo {volume} {304}},\ \bibinfo {pages}
  {229} (\bibinfo {year} {1998})}\BibitemShut {NoStop}%
\bibitem [{\citenamefont {Du}\ \emph {et~al.}(2017)\citenamefont {Du},
  \citenamefont {Fan}, \citenamefont {Zhang},\ and\ \citenamefont
  {Wu}}]{du2017synchronization}%
  \BibitemOpen
  \bibfield  {author} {\bibinfo {author} {\bibfnamefont {L.}~\bibnamefont
  {Du}}, \bibinfo {author} {\bibfnamefont {C.-H.}\ \bibnamefont {Fan}},
  \bibinfo {author} {\bibfnamefont {H.-X.}\ \bibnamefont {Zhang}}, \ and\
  \bibinfo {author} {\bibfnamefont {J.-H.}\ \bibnamefont {Wu}},\ }\href@noop {}
  {\bibfield  {journal} {\bibinfo  {journal} {Scientific reports}\ }\textbf
  {\bibinfo {volume} {7}},\ \bibinfo {pages} {15834} (\bibinfo {year}
  {2017})}\BibitemShut {NoStop}%
\bibitem [{\citenamefont {Johansson}\ \emph {et~al.}(2012)\citenamefont
  {Johansson}, \citenamefont {Nation},\ and\ \citenamefont
  {Nori}}]{johansson2012qutip}%
  \BibitemOpen
  \bibfield  {author} {\bibinfo {author} {\bibfnamefont {J.}~\bibnamefont
  {Johansson}}, \bibinfo {author} {\bibfnamefont {P.}~\bibnamefont {Nation}}, \
  and\ \bibinfo {author} {\bibfnamefont {F.}~\bibnamefont {Nori}},\ }\href@noop
  {} {\bibfield  {journal} {\bibinfo  {journal} {Computer Physics
  Communications}\ }\textbf {\bibinfo {volume} {183}},\ \bibinfo {pages} {1760}
  (\bibinfo {year} {2012})}\BibitemShut {NoStop}%
\bibitem [{\citenamefont {Giovannetti}\ and\ \citenamefont
  {Vitali}(2001)}]{giovannetti2001phase}%
  \BibitemOpen
  \bibfield  {author} {\bibinfo {author} {\bibfnamefont {V.}~\bibnamefont
  {Giovannetti}}\ and\ \bibinfo {author} {\bibfnamefont {D.}~\bibnamefont
  {Vitali}},\ }\href@noop {} {\bibfield  {journal} {\bibinfo  {journal}
  {Physical Review A}\ }\textbf {\bibinfo {volume} {63}},\ \bibinfo {pages}
  {023812} (\bibinfo {year} {2001})}\BibitemShut {NoStop}%
\bibitem [{\citenamefont {Manzano}\ \emph {et~al.}(2013)\citenamefont
  {Manzano}, \citenamefont {Galve}, \citenamefont {Giorgi}, \citenamefont
  {Hern{\'a}ndez-Garc{\'\i}a},\ and\ \citenamefont
  {Zambrini}}]{manzano2013synchronization}%
  \BibitemOpen
  \bibfield  {author} {\bibinfo {author} {\bibfnamefont {G.}~\bibnamefont
  {Manzano}}, \bibinfo {author} {\bibfnamefont {F.}~\bibnamefont {Galve}},
  \bibinfo {author} {\bibfnamefont {G.~L.}\ \bibnamefont {Giorgi}}, \bibinfo
  {author} {\bibfnamefont {E.}~\bibnamefont {Hern{\'a}ndez-Garc{\'\i}a}}, \
  and\ \bibinfo {author} {\bibfnamefont {R.}~\bibnamefont {Zambrini}},\
  }\href@noop {} {\bibfield  {journal} {\bibinfo  {journal} {Scientific
  reports}\ }\textbf {\bibinfo {volume} {3}},\ \bibinfo {pages} {1439}
  (\bibinfo {year} {2013})}\BibitemShut {NoStop}%
\bibitem [{\citenamefont {Giorgi}\ \emph {et~al.}(2012)\citenamefont {Giorgi},
  \citenamefont {Galve}, \citenamefont {Manzano}, \citenamefont {Colet},\ and\
  \citenamefont {Zambrini}}]{giorgi2012quantum}%
  \BibitemOpen
  \bibfield  {author} {\bibinfo {author} {\bibfnamefont {G.~L.}\ \bibnamefont
  {Giorgi}}, \bibinfo {author} {\bibfnamefont {F.}~\bibnamefont {Galve}},
  \bibinfo {author} {\bibfnamefont {G.}~\bibnamefont {Manzano}}, \bibinfo
  {author} {\bibfnamefont {P.}~\bibnamefont {Colet}}, \ and\ \bibinfo {author}
  {\bibfnamefont {R.}~\bibnamefont {Zambrini}},\ }\href@noop {} {\bibfield
  {journal} {\bibinfo  {journal} {Physical Review A}\ }\textbf {\bibinfo
  {volume} {85}},\ \bibinfo {pages} {052101} (\bibinfo {year}
  {2012})}\BibitemShut {NoStop}%
\bibitem [{\citenamefont {Giorgi}\ \emph {et~al.}(2013)\citenamefont {Giorgi},
  \citenamefont {Plastina}, \citenamefont {Francica},\ and\ \citenamefont
  {Zambrini}}]{giorgi2013spontaneous}%
  \BibitemOpen
  \bibfield  {author} {\bibinfo {author} {\bibfnamefont {G.~L.}\ \bibnamefont
  {Giorgi}}, \bibinfo {author} {\bibfnamefont {F.}~\bibnamefont {Plastina}},
  \bibinfo {author} {\bibfnamefont {G.}~\bibnamefont {Francica}}, \ and\
  \bibinfo {author} {\bibfnamefont {R.}~\bibnamefont {Zambrini}},\ }\href@noop
  {} {\bibfield  {journal} {\bibinfo  {journal} {Physical Review A}\ }\textbf
  {\bibinfo {volume} {88}},\ \bibinfo {pages} {042115} (\bibinfo {year}
  {2013})}\BibitemShut {NoStop}%
\bibitem [{\citenamefont {Giovannetti}\ \emph {et~al.}(2002)\citenamefont
  {Giovannetti}, \citenamefont {Lloyd},\ and\ \citenamefont
  {Maccone}}]{giovannetti2002positioning}%
  \BibitemOpen
  \bibfield  {author} {\bibinfo {author} {\bibfnamefont {V.}~\bibnamefont
  {Giovannetti}}, \bibinfo {author} {\bibfnamefont {S.}~\bibnamefont {Lloyd}},
  \ and\ \bibinfo {author} {\bibfnamefont {L.}~\bibnamefont {Maccone}},\
  }\href@noop {} {\bibfield  {journal} {\bibinfo  {journal} {Physical Review
  A}\ }\textbf {\bibinfo {volume} {65}},\ \bibinfo {pages} {022309} (\bibinfo
  {year} {2002})}\BibitemShut {NoStop}%
\bibitem [{\citenamefont {Lee}\ \emph {et~al.}(2014)\citenamefont {Lee},
  \citenamefont {Chan},\ and\ \citenamefont {Wang}}]{lee2014entanglement}%
  \BibitemOpen
  \bibfield  {author} {\bibinfo {author} {\bibfnamefont {T.~E.}\ \bibnamefont
  {Lee}}, \bibinfo {author} {\bibfnamefont {C.-K.}\ \bibnamefont {Chan}}, \
  and\ \bibinfo {author} {\bibfnamefont {S.}~\bibnamefont {Wang}},\ }\href@noop
  {} {\bibfield  {journal} {\bibinfo  {journal} {Physical Review E}\ }\textbf
  {\bibinfo {volume} {89}},\ \bibinfo {pages} {022913} (\bibinfo {year}
  {2014})}\BibitemShut {NoStop}%
\bibitem [{\citenamefont {Zhirov}\ and\ \citenamefont
  {Shepelyansky}(2009)}]{zhirov2009quantum}%
  \BibitemOpen
  \bibfield  {author} {\bibinfo {author} {\bibfnamefont {O.}~\bibnamefont
  {Zhirov}}\ and\ \bibinfo {author} {\bibfnamefont {D.}~\bibnamefont
  {Shepelyansky}},\ }\href@noop {} {\bibfield  {journal} {\bibinfo  {journal}
  {Physical Review B}\ }\textbf {\bibinfo {volume} {80}},\ \bibinfo {pages}
  {014519} (\bibinfo {year} {2009})}\BibitemShut {NoStop}%
\bibitem [{\citenamefont {Roulet}\ and\ \citenamefont
  {Bruder}(2018)}]{roulet2018quantum}%
  \BibitemOpen
  \bibfield  {author} {\bibinfo {author} {\bibfnamefont {A.}~\bibnamefont
  {Roulet}}\ and\ \bibinfo {author} {\bibfnamefont {C.}~\bibnamefont
  {Bruder}},\ }\href@noop {} {\bibfield  {journal} {\bibinfo  {journal}
  {Physical review letters}\ }\textbf {\bibinfo {volume} {121}},\ \bibinfo
  {pages} {063601} (\bibinfo {year} {2018})}\BibitemShut {NoStop}%
\bibitem [{\citenamefont {Ameri}\ \emph {et~al.}(2015)\citenamefont {Ameri},
  \citenamefont {Eghbali-Arani}, \citenamefont {Mari}, \citenamefont {Farace},
  \citenamefont {Kheirandish}, \citenamefont {Giovannetti},\ and\ \citenamefont
  {Fazio}}]{ameri2015mutual}%
  \BibitemOpen
  \bibfield  {author} {\bibinfo {author} {\bibfnamefont {V.}~\bibnamefont
  {Ameri}}, \bibinfo {author} {\bibfnamefont {M.}~\bibnamefont
  {Eghbali-Arani}}, \bibinfo {author} {\bibfnamefont {A.}~\bibnamefont {Mari}},
  \bibinfo {author} {\bibfnamefont {A.}~\bibnamefont {Farace}}, \bibinfo
  {author} {\bibfnamefont {F.}~\bibnamefont {Kheirandish}}, \bibinfo {author}
  {\bibfnamefont {V.}~\bibnamefont {Giovannetti}}, \ and\ \bibinfo {author}
  {\bibfnamefont {R.}~\bibnamefont {Fazio}},\ }\href@noop {} {\bibfield
  {journal} {\bibinfo  {journal} {Physical Review A}\ }\textbf {\bibinfo
  {volume} {91}},\ \bibinfo {pages} {012301} (\bibinfo {year}
  {2015})}\BibitemShut {NoStop}%
\end{thebibliography}%
\bibliographystyle{apsrev4-1}

\end{document}